\documentstyle[prl,aps,tighten,preprint]{revtex}
\begin{document}
\draft
\title{Current Algebra in the Path Integral framework}
\author{V. Cardenas\thanks{%
E-mail: vcardena@lauca.usach.cl}, S. Lepe\thanks{%
E-mail: slepe@lauca.usach.cl} and J. Saavedra\thanks{%
E-mail: jsaavedr@lauca.usach.cl}}
\address{Departamento de Fisica, Universidad de Santiago de Chile,\\
Casilla 307, Santiago 2, Chile }
\maketitle

\begin{abstract}
In this letter we describe an approach to the current algebra based in the
Path Integral formalism. We use this method for abelian and non-abelian
quantum field theories in 1+1 and 2+1 dimensions and the correct expressions
are obtained. Our results show the independence of the regularization of the
current algebras.
\end{abstract}


Abelian bosonization and current algebras play an important role in the
description of two-dimensional quantum field theories as non-perturbative
methods and they are an important ingredient in order to show the
equivalence between different (two-dimensional) models \cite
{cole,mandel,gambo} (for a complete review of the most important references
in the field see \cite{review}).

The non-abelian extension of the bosonization is, however, a more technical
problem that was solved in \cite{witt}. Essentially, the solution given by
Witten for $N$ free fermionic fields was to show the equivalence with a
Wess-Zumino-Witten \cite{wznw} theory with the current algebra describing a $%
SU(N)$ Kac-Moody algebra.

In the abelian or non-abelian bosonization, the current commutators are
normally computed using a point splitting regularization plus the
Bjorken-Johnson-Low (BJL) limit, in order to have an equal-time commutator.
Although it seems a technical point, the computation of the current-current
commutator using different regularizations could shed some light on the
independence of the regularization of the current algebra\cite{affleck}.

The purpose of this paper is to present an explicit calculation of the
current algebra in two and three dimensions based in the path integral
approach. This procedure allows translating the definition of the product of
two operators at the same point, to a regularization of a functional
determinant where many other regularizations are available.

In order to compute the current algebra, let us start considering a massless
fermion in 1+1 dimensions coupled to a gauge field $A_\mu $ 
\begin{equation}
L=\bar{\psi}iD\hspace {-0.6 em}\slash\hspace {0.15 em}\psi ,  \label{lagr}
\end{equation}
where $D_\mu =\partial _\mu +A_\mu $.

The gauge field $A_\mu $ is an external auxiliary field that can be set
equal to zero at the end of the calculation.

Thus, from the euclidean partition function

\begin{equation}
Z\left[ A\right] =\int D\bar{\psi}D\psi \exp \left[ -\int d^2x\bar{\psi}iD%
\hspace {-0.6 em}\slash\hspace {0.15 em}\psi \right] ,  \label{e2}
\end{equation}
one can compute the current-current correlator by means 
\begin{equation}
\left\langle J^\sigma (z)J^\rho (w)\right\rangle =\frac{\delta ^2Z}{\delta
A_\sigma (z)\delta A_\rho (w)}\mid _{A=0}.  \label{meth}
\end{equation}
However, as $Z=\det (iD\hspace {-0.6 em}\slash\hspace {0.15 em}),$ the
calculation of (\ref{meth}) requires a regularization prescription for the
fermionic determinant. This is the main advantage of our method, because one
could use different regularizations for the determinant and, as a
consequence, the commutator 
\begin{equation}
\left\langle \lbrack J^\sigma (z),J^\rho (w)]\right\rangle \equiv (\frac{%
\delta ^2Z}{\delta A_\sigma (z)\delta A_\rho (w)}-\frac{\delta ^2Z}{\delta
A_\rho (w)\delta A_\sigma (z)})\mid _{A=0},  \label{conmut}
\end{equation}
showing the universability of the current algebra.

There is, however, a delicate point; the commutator must be computed at
equal times implying one aditional regularization, namely the BJL limit \cite
{bjl,tsv}. Below we show explicitly these calculations for several examples.

In order to prove our previous assertions, let us start considering the 1+1
abelian case where the fermionic determinant is exact.

In the 1+1 dimensional case considered above, the fermionic determinant
becomes 
\begin{equation}
Z\left[ A\right] =\det (iD\hspace {-0.6 em}\slash\hspace {0.15 em})=\exp
\left[ -\frac 1{4\pi }\int d^2xF_{\mu \nu }\Box ^{-1}F^{\mu \nu }\right] ,
\label{e3}
\end{equation}
where the determinant have been regularized using the $\zeta -$function
method.

From (\ref{e3}) it is easy to obtain 
\begin{equation}
\frac{\delta ^2Z}{\delta A_\sigma (z)\delta A_\rho (w)}=-\frac{Z[A]}\pi \int 
\frac{d^2p}{\left( 2\pi \right) ^2}\left\{ \delta ^{\rho \sigma }-\frac{%
\left( p^\rho \cdot p^\sigma \right) }{p^2}\right\} \exp \left( -ip\left(
w-z\right) \right) ,  \label{e32}
\end{equation}
where the terms proportional to $A$ have been omitted. By interchanging ($%
\rho $, $\sigma )$ with ($w,$ $z),$ one finds the covariant commutator 
\begin{equation}
\left\langle \left[ J^\rho (z),J^\sigma (w)\right] \right\rangle _{A=0}=%
\frac{2i}\pi \int \frac{d^2p}{\left( 2\pi \right) ^2}\left\{ \delta ^{\rho
\sigma }-\frac{\left( p^\rho \cdot p^\sigma \right) }{p^2}\right\} \sin
\left( p\left( w-z\right) \right) .  \label{e34}
\end{equation}
In order to satisfy the microcausality principle one must, additionally,
impose the limit $t\rightarrow t^{\prime }.$ For the commutator $%
\left\langle [A(x),B(y)]\right\rangle $ this limit means 
\begin{equation}
\left\langle \lbrack A(x),B(y)]\right\rangle =\lim_{\tau \rightarrow
+0}(\left\langle A(\tau ,x)B(0,y)\right\rangle -\left\langle A(-\tau
,x)B(0,y)\right\rangle ).  \label{bjl}
\end{equation}
Using this fact and equation (\ref{e32}), we find

\begin{equation}
\left\langle \left[ J_0(z),J_0(w)\right] \right\rangle =\left\langle \left[
J_1(z),J_1(w)\right] \right\rangle =0,  \label{ex0}
\end{equation}

\begin{equation}
\left\langle \left[ J_0(z),J_1(w)\right] \right\rangle =\frac{-2i}\pi \int 
\frac{d^2p}{\left( 2\pi \right) ^2}\frac{p_0p_1}{p^2}\lim_{\tau \rightarrow
+0}\left\{ \sin (p_0\tau )\right\} \exp (-ip_1(z-w)),  \label{ex1}
\end{equation}
where $\tau =t-t^{\prime }$.

The integrals in (\ref{ex1}), can be evaluated using the well known trick $%
a^{-1}=\int_0^\infty dx\exp (-ax)$ and after integrating in $p_0$ we obtain 
\begin{equation}
\left\langle \left[ J_0(z),J_1(w)\right] \right\rangle =\frac{-i}\pi \int 
\frac{dp_1}{\left( 2\pi \right) ^2}p_1\exp (-ip_1(z-w))\sqrt{\pi }\lim_{\tau
\rightarrow +0}\tau (4p_1^2/\tau ^2)^{1/4}K_{\frac 12}(\sqrt{\tau ^2p_1^2}).
\label{ex2}
\end{equation}
The limit $\tau \rightarrow +0$ is computed using 
\[
\begin{array}{cccc}
K_{\frac 12}(z)\rightarrow \frac 12\sqrt{2\pi /z} &  & \text{when} & z\ll 1,
\end{array}
\]
and the commutator (\ref{ex2}) becomes 
\begin{eqnarray}
\left\langle \left[ J_0(z),J_1(w)\right] \right\rangle &=&\frac{-i}{2\pi }%
\int \frac{dp_1}{\left( 2\pi \right) }p_1\exp (-ip_1(z-w))  \label{alg1} \\
&=&-\frac 1{2\pi }\delta ^{^{\prime }}\left( z-w\right) ,  \nonumber
\end{eqnarray}
that is the expected Schwinger term \cite{schw}.

One can extend the above result for the 2+1 abelian case where the fermionic
determinant in powers of $1/m$ is also available. Indeed, using a
Pauli-Villars regulator the partition function becomes 
\begin{equation}
Z\left[ A\right] =\exp \left[ -\frac m{16\pi \left| m\right| }\int
d^3x\epsilon ^{\mu \alpha \beta }A_\mu F_{\alpha \beta }+O\left( \frac 1m%
\right) \right] .  \label{e7}
\end{equation}
Using again equation (\ref{meth}), one finds 
\begin{equation}
\left\langle J^\rho (z)J^\sigma (w)\right\rangle =-\frac 1{4\pi }\frac m{%
\left| m\right| }\epsilon ^{\rho \sigma \mu }\partial _\mu \delta
^{(3)}(z-w),  \label{e8}
\end{equation}
and the current-current commutator reads

\begin{equation}
\left\langle \left[ J^\rho (z),J^\sigma (w)\right] \right\rangle _0=-\frac 1{%
2\pi }\frac m{\left| m\right| }\epsilon ^{\mu \rho \sigma }\partial _\mu
\delta ^{(3)}(z-w).  \label{e16}
\end{equation}
Using the identity 
\[
\delta ^{(3)}(z-w)=\int \frac{d^3p}{(2\pi )^3}\exp (-ip(z-w)), 
\]
the commutator (\ref{e16}) after taking the BJL limit becomes 
\[
\left\langle \left[ J^\rho ({\bf z}),J^\sigma ({\bf w})\right] \right\rangle
\mid _{t=t^{\prime }}=-\frac 1{4\pi }\frac m{\left| m\right| }\epsilon
^{\rho \sigma \mu }\int \frac{d^3p}{(2\pi )^3}(-ip_\mu )\lim_{\tau
\rightarrow +0}-2i\sin (p_0\tau )\exp (-i{\bf p}({\bf z}-{\bf w})). 
\]
To solve these integrals a converging factor $\exp (-\alpha p_0)$ is added
to the integrand, and we take the limit $\alpha \rightarrow 0$ at the end.
We explicity find 
\[
\left\langle \left[ J^0({\bf z}),J^1({\bf w})\right] \right\rangle =-\frac 1{%
4\pi }\frac m{\left| m\right| }\partial _2\delta ^{(2)}({\bf z}-{\bf w}%
)\lim_{\tau \rightarrow +0}\frac 1\tau ,\qquad 
\]
\begin{equation}
\left\langle \left[ J^0({\bf z}),J^2({\bf w})\right] \right\rangle =-\frac 1{%
4\pi }\frac m{\left| m\right| }\partial _1\delta ^{(2)}({\bf z}-{\bf w}%
)\lim_{\tau \rightarrow +0}\frac 1\tau ,  \label{alg2}
\end{equation}
\[
\left\langle \left[ J^1({\bf z}),J^2({\bf w})\right] \right\rangle =-\frac 1{%
6\pi }\frac m{\left| m\right| }\delta ^{(2)}({\bf z}-{\bf w})\lim_{\tau
\rightarrow +0}\frac 1{\tau ^2},\ \qquad 
\]
where, in the last expression, $\delta (\tau )$ was written using the
representation $\delta (\tau )=\pi ^{-1}\lim_{\alpha \rightarrow 0}\alpha
/(\tau ^2+\alpha ^2).$

The current algebra (\ref{alg2}) could be used as starting point in
applications to condensed matter physics as was done in \cite{wen}. However,
(\ref{alg2}) depends on the regulator and we do not find agreement with \cite
{wen}. The same algebra (\ref{alg2}) was derived also in \cite{shapos} using
differents methods.

The extensions to the non-abelian case is performed as follows; Firstly, let
us consider the 1+1 dimensional case. One can proceed along the same lines
because the fermionic determinant is proportional to $e^{-S_{WZW}},$ where $%
S_{WZW}$ is the Wess-Zumino-Witten effective action \cite{wznw,polywe} given
by 
\begin{equation}
S_{WZW}=\frac 12\int d^2xtr(\partial _\mu g^{-1}\partial ^\mu g)+\frac i{%
8\pi }\int d^3\xi \epsilon ^{abc}tr(g^{-1}\partial _agg^{-1}\partial
_bgg^{-1}\partial _cg),  \label{awzw}
\end{equation}
where g is the $SU(N)$ matrix. The auxiliar field ${\bf A}^\mu {\bf =}%
A_i^\mu \sigma _i$ written in light cone coordinates is represented in terms
of $g$ as ${\bf A}_{+}=g^{-1}\partial _{+}g,$ and in the light cone gauge $%
{\bf A}_{-}=0$. Thus, the partition function is 
\begin{equation}
Z\left[ g\right] =\exp [-S_{WZW}],  \label{e4}
\end{equation}
where locally, (\ref{awzw}) was written as 
\[
S_{WZW}=\int_0^1d\alpha (1-\alpha )\int d^2xtr(\partial _{-}\phi e^{-\alpha
\phi }\partial _{+}\phi e^{\alpha \phi }), 
\]
with $g=e^{i\phi }$ with $\phi $ an antihermitian matrix \cite{polywe}.

Then we can write $\partial _{+}\phi =-i{\bf A}_{+}$ and $\partial _{-}\phi
=-i\int dy^{+}\partial _{-}{\bf A}_{+}(y^{+},x^{-})$ and using (\ref{meth})
we obtain 
\[
\frac{\delta ^2Z\left[ A\right] }{\delta A_{+}^a(z)\delta A_{+}^b(w)}\mid
_{A=0}=-\frac 12\int dx_{+}dx_{-}tr(\frac{\delta (\partial _{-}\phi )}{%
\delta A_{+}^a(z)}\frac{\delta (\partial _{+}\phi )}{\delta A_{+}^b(w)}), 
\]
\[
\hspace{0.9cm}=-\frac 12\delta _{ab}\delta ^{\prime }(w_{-}-z_{-}), 
\]
where we imposed $\phi =0,$ so that $g={\bf I}$ and ${\bf A}_{+}=0$ (here
the normalization $tr(\sigma _a\sigma _b)=2\delta _{ab}$ has been used).

In analogy with the previous case we obtain

\begin{equation}
\left\langle \left[ J_a(x_{-}),J_b(y_{-})\right] \right\rangle
=if_{abc}J_c(x_{-})\delta (x_{-}-y_{-})+\delta _{ab}\delta ^{\prime
}(x_{-}-y_{-}),  \label{e5}
\end{equation}
where (\ref{e5}) is the $SU(N)$ Kac-Moody algebra of level $k=1$. We should
note that the use of light cone coordinates and the light cone gauge
simultaneously allowed us to obtain expression (\ref{e5}) at equal times
without taking the BJL limit that was needed before (see \cite{affleck}).

The non-abelian extension to 2+1 dimensions, can be performed as follows.
Start from the well-known expression for the partition function 
\begin{equation}
Z[A]=\exp \left[ -\frac 1{8\pi }\frac m{\left| m\right| }tr\int d^3x\left( 
{\bf A}_\mu {}^{*}{\bf F}^\mu -\frac i3\varepsilon ^{\mu \alpha \beta }{\bf A%
}_\mu {\bf A}_\alpha {\bf A}_\beta \right) +O\left( \frac 1m\right) \right] .
\label{let}
\end{equation}
Since we must take ${\bf A}_\mu =0$ at the end of the calculation, the
second term in the argument of (\ref{let}) does not contribute. Then, we
work with an action similar to (\ref{e7}) but written in terms of the
non-abelian field ${\bf A}_\mu $ and the stress tensor ${\bf F}_{\mu \nu
}=\partial _\mu {\bf A}_\nu -\partial _\nu {\bf A}_\mu +\left[ {\bf A}_\mu ,%
{\bf A}_\nu \right] $ . Following the same lines describe above, we find the
Kac-Moody algebra 
\begin{equation}
\left[ J_a^\rho (x),J_b^\sigma (y)\right] =if_{abc}J_c^\sigma (x)\gamma
^0\gamma ^\rho \delta (x-y)+i\frac{\delta _{ab}}{4\pi }\frac m{\left|
m\right| }\epsilon ^{\rho \sigma \mu }\partial _\mu \delta ^{(3)}(x-y).
\label{ue}
\end{equation}
The BJL limit is taken in analogy with (\ref{e8}).

In conclusion we have described an approach based in the path integral
formalism, where the current algebra can be computed for quantum field
theories in 1+1 and 2+1 dimensions. This approach could be considered as an
alternative method which permits obtaining the current algebra in different
regularizations.

The authors would like to thank Dr. J. Gamboa for suggesting this problem to
us and guidance throughout this work. We would also like to thank J. Zanelli
for the critical reading of the manuscript.

\end{document}